\documentclass[manuscript]{aastex}

\usepackage{rotating}
\usepackage{tikz}
\usepackage{amsmath}

\slugcomment{The Astrophysical Journal Letters}

\shorttitle{Fragmenting Comet 332P/Ikeya-Murakami}
\shortauthors{Jewitt et al.}

\begin{document}

\title{Fragmentation Kinematics in  Comet 332P/Ikeya-Murakami}


\author{David Jewitt$^{1,2}$,  Max Mutchler$^3$, Harold Weaver$^4$, Man-To Hui$^1$, Jessica Agarwal$^5$, Masateru Ishiguro$^6$, Jan Kleyna$^7$, Jing Li$^{1}$, Karen Meech$^7$, Marco Micheli$^8$,  Richard Wainscoat$^7$,  and  Robert Weryk$^7$
}

\affil{$^1$Department of Earth, Planetary and Space Sciences,
UCLA, 595 Charles Young Drive East, Los Angeles, CA 90095-1567\\
$^2$Department of Physics and Astronomy, University of California at Los Angeles, 430 Portola Plaza, Box 951547, Los Angeles, CA 90095-1547\\
$^3$ Space Telescope Science Institute, 3700 San Martin Drive, Baltimore, MD 21218 \\
$^4$ The Johns Hopkins University Applied Physics Laboratory, 11100 Johns Hopkins Road, Laurel, Maryland 20723  \\
$^5$ Max Planck Institute for Solar System Research, Justus-von-Liebig-Weg 3, 37077 G\"ottingen, Germany \\
$^6$ Department of Physics and Astronomy, Seoul National University,
Gwanak, Seoul 151-742, Republic of Korea \\
$^7$ Institute for Astronomy, University of Hawaii, 2680 Woodlawn Drive, Honolulu, HI 96822 \\
$^8$ SSA NEO Coordination Centre, European Space Agency, 00044 Frascati, RM, Italy \\
}

\email{jewitt@ucla.edu}


\begin{abstract}
We present initial time-resolved observations of the split comet 332P/Ikeya-Murakami taken using the Hubble Space Telescope.  Our images reveal a dust-bathed cluster of fragments receding from their parent nucleus at projected speeds in the range 0.06 to 3.5 m s$^{-1}$ from which we estimate ejection times from October to December 2015.  The number of fragments with effective radii $\gtrsim$20 m follows a differential power law with index $\gamma$ = -3.6$\pm$0.6, while smaller fragments are less abundant than expected from an extrapolation of this power-law.  We argue that, in addition to losses due to observational selection, torques from anisotropic outgassing are capable of destroying the small fragments by driving them quickly to rotational instability.   Specifically, the spin-up times of  fragments $\lesssim$20 m in radius are shorter than the time elapsed since ejection from the parent nucleus.  The effective radius of the parent nucleus is $r_e \le$ 275 m (geometric albedo 0.04 assumed).  This is about seven times smaller than previous estimates  and  results in a nucleus mass at least 300 times smaller than previously thought.   The  mass in solid pieces, $2\times10^9$ kg, is about 4\% of the mass of the parent nucleus.  As a result of its small size, the  parent nucleus also has a short spin-up time.  Brightness variations in time-resolved nucleus photometry are consistent with rotational instability playing a role in the release of fragments.

\end{abstract}


\section{Introduction}

Short-period comet 332P/Ikeya-Murakami (formerly P/2010 V1, hereafter  ``332P''), was discovered visually at heliocentric distance $r_H$ = 1.601 AU on UT 2010 November 2 (one month after perihelion on UT 2010 October 13; Nakano and Ikeya 2010a). Its orbit has semimajor axis $a$ = 3.088 AU, eccentricity $e$ = 0.491, inclination $i$ = 9.4\degr~and perihelion distance $q$ = 1.573 AU.  332P is a short-period comet (orbital period 5.43 yr), likely to have survived a $\sim$10 Myr  journey to the inner solar system (Tiscareno and Malhotra 2003) following 4.5 Gyr spent in the Kuiper belt.  

Subsequent observations over three months showed 332P  to fade steadily at about 6\% per day (Ishiguro et al.~2014).  At its peak, the dust mass in the coma was estimated at $\sim$5$\times$10$^8$ kg, corresponding to $>2\times$10$^{-5}$ of the mass of the nucleus (taken by Ishiguro et al.~as a sphere of $<$1.85 km radius and density $\rho$ = 1000 kg m$^{-3}$).  No fragmentation of the nucleus was reported.  The morphology, the steady fading and a non-detection of the comet on UT 2010 November 1 (Nakano and Ikeya 2010b) suggest that 332P was discovered because of its photometric outburst. Presumably, it went undiscovered before the outburst as a result of  low or negligible outgassing activity.  A similarity to the archetypal outbursting comet 17P/Holmes (c.f.~Hsieh et al.~2010, Li et al.~2011) was duly noted (Ishiguro et al.~2014).  In both comets, runaway crystallization of amorphous ice was implicated as a possible driver of the activity.

On UT 2015 December 31 (three months before the subsequent perihelion on UT 2016 March 14), the parent nucleus (now known as 332P-C) was reported to be accompanied by a companion (332P-A), leading to the realization that 332P had split (Weryk et al.~2016).  Kleyna et al.~(2016) estimate that 332P-A split from 332P-C in early 2014 (uncertainty of six months) while Sekanina (2016) reported UT 2012 December 1$\pm$31 days (when $r_H$ = 4.44 AU).  Continued observations with ground-based telescopes revealed additional fragments (Kleyna et al.~2016), but interpretation  of these observations is made difficult by the limited resolution and depth of the reported ground-based data. 

We secured target-of-opportunity observing time on the Hubble Space Telescope (HST) in order to examine 332P at the highest angular resolution.    While HST has been used before to examine fragmenting comets (Weaver et al.~1995, 2001), this is the first time that observations have been secured with a cadence sufficient to study the fragment kinematics.  Here, we report initial measurements from three days in 2016 January and from a sequence of images taken in April to examine short-term variability.  

\section{Observations}
Observations were obtained using the HST under programs GO 14474 and 14498.  Within each orbit, we obtained five consecutive integrations of 420 s with the WFC3 camera (Dressel 2015). To obtain maximum sensitivity, we employed  the F350LP filter, which has a central wavelength near $\sim$6230\AA~and a full width at half maximum of $\sim$4758 \AA~when observing a source with a sun-like spectrum.  We  dithered the exposures to mitigate the effects from bad pixels, cosmic rays, and the inter-chip gap.  The earliest possible observations were secured on UT 2016 January 26, 27 and 28.

The appearance of 332P is shown in Figure (\ref{image}).  Arcs and streaks in the figure are residual images of field stars and galaxies trailed by parallactic motion of the telescope. The top panel shows, in addition to the parent nucleus ``332P-C" and the bright companion identified by Weryk et al.~(2016) (called ``332P-A"),  a cluster of fragments located to the west of ``332P-C'' and distributed about the A-C axis (which is also the direction of the projected orbit).  With few exceptions, the cluster fragments cannot be unambiguously associated with components already identified by ground-based observers, because of blending and sensitivity differences and also because of rapid evolution of the fragments (e.g.~Kleyna et al.~2016).  Therefore, we employ our own labels, given as lower-case letters in Figure (\ref{image}). Close inspection of data from January 26, 27 and 28 shows that the fragments move and evolve both photometrically and, in some cases, morphologically.  We base the present study on fragments which could be identified and cross-linked over the three days of observation.  Additional fragments, appearing in just one or two of the three epochs of observation, will be the study of a future paper, as will an attempt to link the fragments seen in January to those detected in later months.

\subsection{Dynamics}
We measured the positions of the fragments using median-combined composite images created for each day of observation.  Most objects were digitally centroided within a 5 pixel wide box but particularly faint and/or blended fragments were centroided by eye.  In all cases, the positional uncertainty is $\lesssim \pm$1 pixel (0.04\arcsec, or about 20 km).  We also determined photometry for each fragment, discussed in the next section.  Relative movements of the fragments are clearly visible from day to day resulting from characteristic velocities of order a few meters per second.    Figure (\ref{v_d_plot}) shows the sky-plane velocity, $v$, measured from the January 26, 27 and 28 data versus the projected distance, $\ell$, from the parent ``332P-C''.  Uncertainties on the data points are mostly smaller than the symbols used to plot the data.  

The simplest and most natural interpretation of the linear velocity vs.~distance plot is that the fragments were ejected simultaneously  with a range of velocities from as small as 0.06 m s$^{-1}$ (fragment w) to as large as 3.5 m s$^{-1}$ (fragment v); the fastest fragments have traveled the greatest distances.  A weighted, least-squares fit to the data, forced to pass through the origin, gives $v = (1.9\pm0.2)\times10^{-7} \ell$, with $\ell$ in meters and $v$ in m s$^{-1}$.   The corresponding time of flight, assuming that the fragments are unaccelerated, is simply $\tau = \ell/v$.  We find $\tau$ = 61$\pm$6 days, corresponding to a single ejection date on UT 2015 November 27$\pm$6,  one month before the discovery of the split nature of the comet (Weryk et al.~2016). However, Figure (\ref{v_d_plot}) shows a significant dispersion of data points around the best-fit line, corresponding to a range of flight times as marked in the figure and shown as a histogram inset.  Flight times from $\sim$40 to $\sim$80 days are indicated, corresponding to ejection dates between about UT 2015 October 19 and December 18 (c.f.~Kleyna et al.~2016).  The spread of ejection times argues against an impulsive (e.g.~impact) origin, as does the earlier ejection of component A and the existence of  more distant (older) components projected outside the HST field of view (Kleyna et al.~2016, Sekanina 2016).

Other interpretations of the linear $v \propto \ell$ relation (Figure \ref{v_d_plot}) are possible.  For example, at least part of each fragment's motion is due to the divergence of Keplerian orbits caused by the ejection velocities. We have neglected the effects of projection into the plane of the sky. The fragments appear to have been released from the parent nucleus over a range of times, not simultaneously, and some fragments could be tertiary products of break-up occurring during flight.  The fragment motions could also be influenced by non-gravitational accelerations due to asymmetric outgassing, although an initial search for this effect has been unsuccessful.  These should scale inversely with object size, imbuing  smaller fragments with larger  velocities in a given time. However, we  find no evidence for  a  relation between fragment brightness (a proxy for size) and speed (but such a relation could be hidden if the brightness does not provide a measure of fragment size, c.f.~Section \ref{sizes}). Non-gravitational acceleration would not necessarily lead to a speed vs.~distance relation of the form observed.   These and other possibilities may be tested by the inclusion of additional data taken in later months.  However, the basic conclusions (that the fragments were ejected recently and with low velocity) are robust. With its age measured in months, the cluster of fragments is clearly the product of an event distinct from the  photometric outburst in 2010 and from the  separation of components 332P-C and 332P-A in late 2012, consistent with cascading fragmentation of the type exhibited by the Kreutz sungrazers (Sekanina 2002).

%


\subsection{Size and Size Distribution}
\label{sizes}
The photometry provides a measure of the sum of the scattering cross-sections of all the particles (dust and nucleus) inside the photometry aperture.  The spatial resolution afforded by HST allows us to reject near-nucleus dust with an order-of-magnitude greater efficiency than is possible in typical ground-based data.  However, the resulting cross-sections must still be interpreted as upper limits to the cross-section of macroscopic bodies in the aperture owing to residual dust contamination.  To minimize contaminating dust we measured each fragment using the smallest photometry aperture (radius 0.2\arcsec, corresponding to 240 km at $r_H$ = 1.64 AU) with background subtraction from a contiguous annulus having outer radius 0.8\arcsec.  The resulting apparent magnitudes, $V$, were converted to absolute magnitudes using 

\begin{equation}
H = V - 5\log_{10}(r_H \Delta) - \beta \alpha
\label{abs}
\end{equation}

\noindent in which $r_H$ and $\Delta$ are the heliocentric and geocentric distances, respectively, and $\beta$ is a measure of the phase darkening at phase angle $\alpha$.  The phase coefficient is unmeasured in 332P; we take $\beta$ = 0.04 magnitudes per degree based on observations of other comets.  Uncertainties in the derived $H$ are dominated by our ignorance of $\beta$, rather than by uncertainties in the photometry (e.g.~a $\beta$ value larger or smaller by 0.01 magnitudes per degree would change $H$ in the January data by $\pm$0.1 magnitudes).  

The absolute magnitude is further interpreted in terms of scattering from an effective cross-section $C_e$ (km$^2$), using

\begin{equation}
C_e = \frac{1.5\times10^6}{p_v} 10^{-0.4 H}
\label{radius}
\end{equation}

\noindent where $p_V$ is the geometric albedo.  We assume $p_V$ = 0.04, compatible with measurements of comets.  The radius of a circle having cross-section $C_e$ is $r_e = (C_e/\pi)^{1/2}$.  The resulting radii are strictly to be interpreted as upper limits to the radii of solid fragments, because of contamination by dust.  Even so, the derived values are remarkably small, ranging from $r_e \sim$ 10 m for the smallest pieces to 275 m for the two brightest, largest components ``C" and ``A".  

The cumulative  distribution of fragment cross-sections is shown in Figure (\ref{N_vs_C}), plotted separately for each of the three days of measurement.  The distribution is consistent with a broken power law, with an inflection at $C_e$ = 1200 m$^2$ (equivalent circular radius $r_e = (C_e/\pi)^{0.5} \sim$ 20 m).   We write the differential distribution of cross-sections as $n(C_e)dC_e = G r_e^{-g} dC_e$, where $G$ and $g$ are constants.  At  $C_e >$ 1200 m$^2$, the slope of the cumulative distribution  is $1-g$ = -1.3$\pm$0.3, giving $g$ = 2.3$\pm$0.3.

If the number of fragments with radii between $r_e$ and $r_e + dr_e$ is written $n(r_e)dr_e = \Gamma r_e^{-\gamma} dr_e$, and if the apparent brightness of a fragment is proportional to $r_e^2$, then the distribution of fragment brightnesses should obey a power law with index $\gamma =2g - 1$.  With $g$ = 2.3$\pm$0.3 for the larger objects, we infer $\gamma$ = 3.6$\pm$0.6.   For comparison, the size distribution of $>$10 m sized boulders measured on the nucleus of 103P/Hartley 2 follows $\gamma = 3.7\pm0.2$ (Pajola et al.~2016) while Ishiguro et al.~(2009) reported $\gamma = 3.34\pm0.05$ in ejected fragments of 73P/Schwassmann-Wachmann 3.    The Kreutz family comets follow $\gamma$ = 3.2 between radii of about 5 m and 35 m (Knight et al.~2010).  (For unknown reasons, boulders on 67P follow a steeper distribution, with  $\gamma$ = 4.6$^{+0.2}_{-0.3}$; Pajola et al.~2015).   The mass in distributions with $\gamma <$ 4  is dominated by the largest (brightest) particles in the distribution, indicating that our observations provide a meaningful estimate of the total mass.

\section{Discussion}

\begin{figure}
\plotone{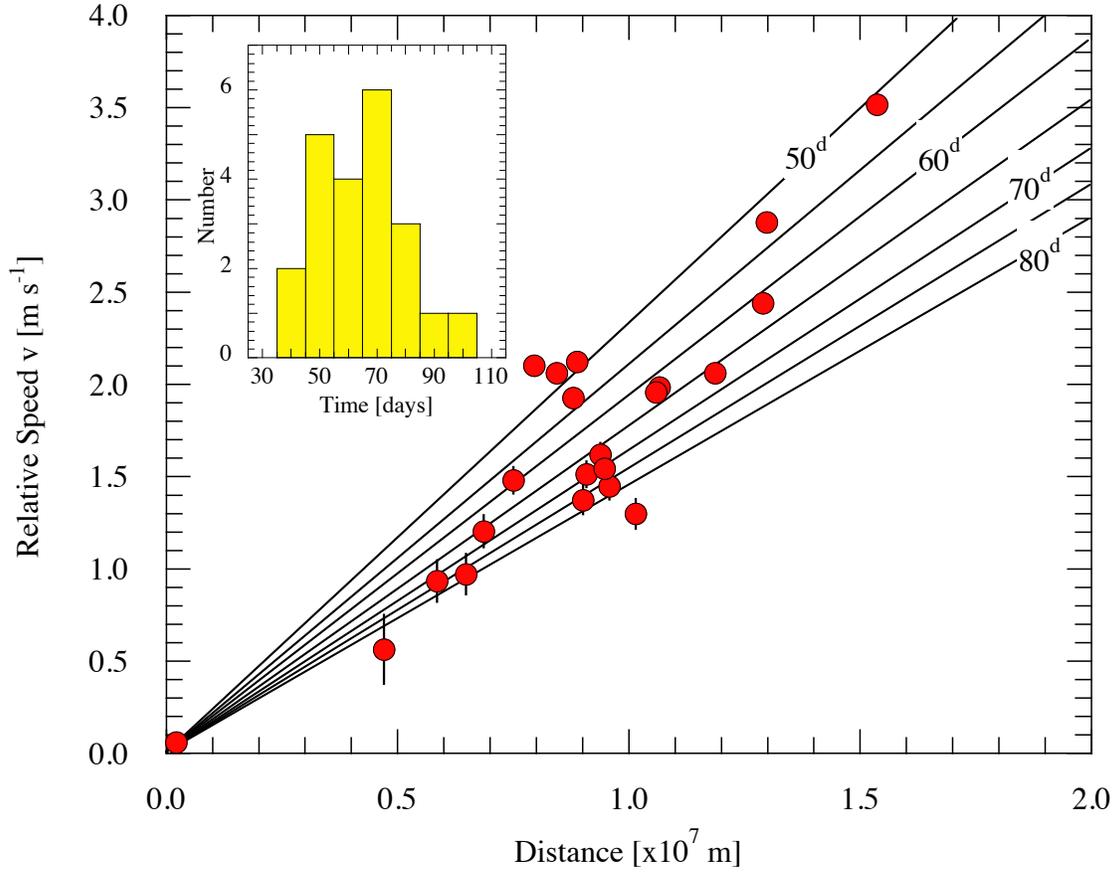}
\caption{Measured fragment velocities  as a function of sky-plane distance, both with respect  to parent body ``C''.  Diagonal lines show different times since ejection, labeled in days.  The inset plots the distribution of travel times.  \label{v_d_plot}}
\end{figure}

\begin{figure}
\epsscale{.80}
\plotone{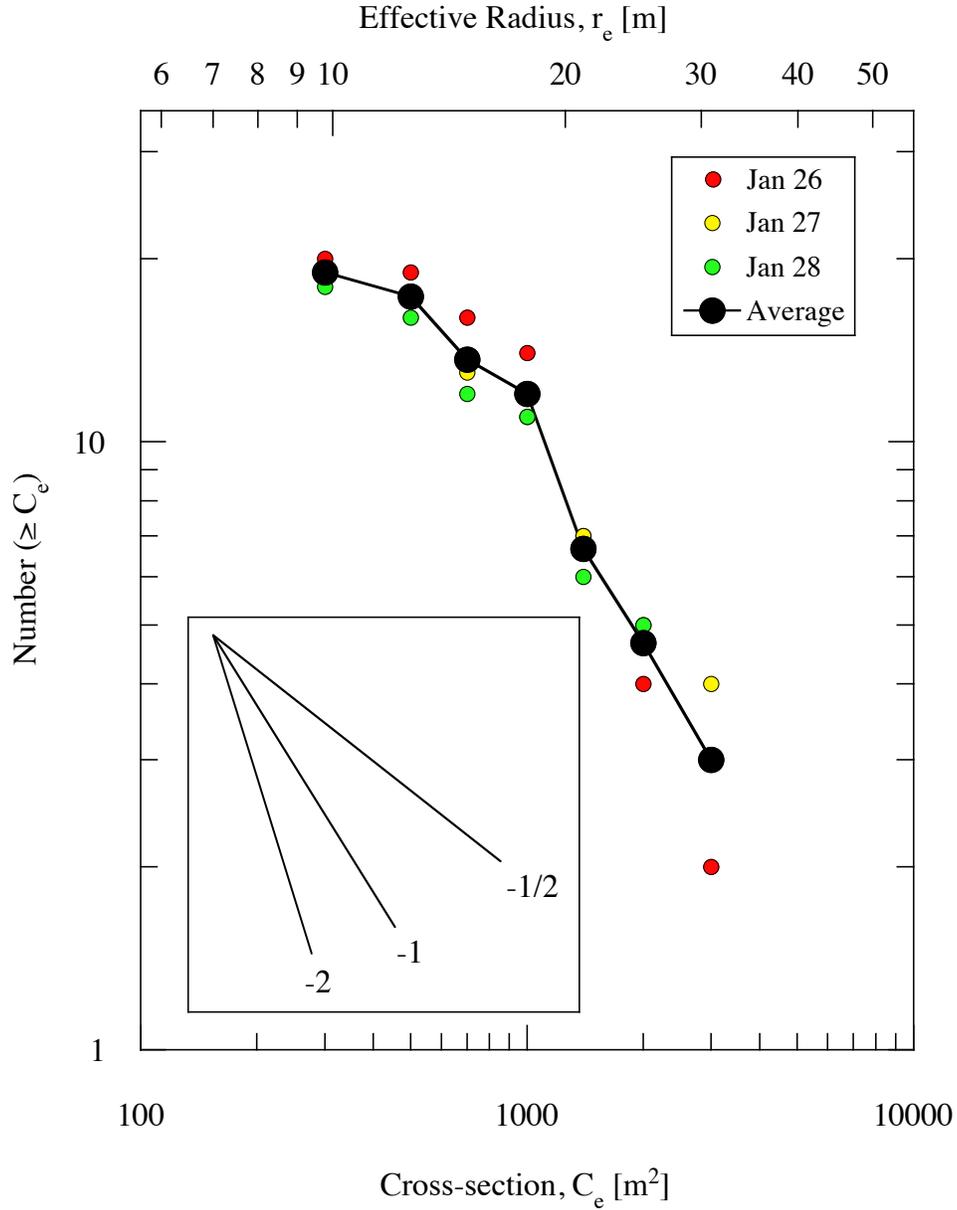}
\caption{Cumulative distribution of fragment cross-sections, $C_e$, computed from Equation (\ref{radius}), (lower axis, in m$^2$) and effective radii given by $r_e = (C_e/\pi)^{1/2}$ (upper axis, in m).  Large black circles show the average counts from UT 2016 January 26, 27 and 28 while small color-coded circles show the individual counts. The inset shows gradients $1-g$ = -1/2, -1 and -2, for reference. \label{N_vs_C}}
\end{figure}

At $r_e \lesssim$ 20 m  the size distribution is more nearly flat, $1-g$ = -0.5$\pm$0.3 ($\gamma$ = 2.0$\pm$0.6).  Part of the flattening may be caused by observational selection, which discriminates against the detection of faint fragments.  In our data (composite images with total integration times 2100 s) a signal-to-noise ratio = 3 is reached on solar-spectrum, point-source targets with V = 28.4, corresponding to $H$ = 27.7 (Equation \ref{abs}), $C_e$ = 300 m$^2$ (Equation \ref{radius}) and $r_e$ = 10 m, considerably smaller than the knee in Figure (\ref{N_vs_C}).  As an additional process, we speculate that rapid destruction of  smaller fragments   also contributes to the flattening of the distribution.  Here, we show that the timescale for spin-up of  fragments  to  centripetal instability by sublimation torques is shorter than the time since ejection of the fragments indicated by their motion (Figure \ref{v_d_plot}), provided $r_e \lesssim$ 20 m.

Anisotropic mass loss from an irregular body produces a torque which can affect the spin.  The e-folding timescale for spin-up to the centripetal limit (beyond which neither gravity nor cohesive forces can maintain the structure) is (Jewitt 1997)

\begin{equation}
\tau_s \sim \frac{\omega \rho r_e^4}{V_{th} k_T (dM/dt)}
\label{spin1}
\end{equation}

\noindent where $\omega = 2\pi/P$ is the initial angular frequency of rotation at period $P$, $\rho$ is the mass density, $r_e$ is the radius of the body, $V_{th}$ is the thermal speed of the sublimated gas, $k_T$ is the dimensionless moment-arm for the torque and $dM/dt$ is the mass loss rate due to sublimation.  We set $dM/dt = \pi k_A r_e^2 f_s$, where $k_A$ is the fraction of the surface in active sublimation and $f_s$ is the specific sublimation rate from the surface.  Then, substituting into Equation (\ref{spin1}) and neglecting constants of order unity, we obtain

\begin{equation}
\tau_s \sim \frac{\rho r_e^2}{V_{th} k_T k_A f_s  P}.
\label{spin2}
\end{equation}

\noindent  We take $P$ = 5 hr, typical of small bodies, $\rho$ = 500 kg m$^{-3}$ (c.f.~Jorda et al.~2016) and $V_{th}$ = 500 m s$^{-1}$ as appropriate for water sublimating at 200 K.  Moment arm, $k_T$, is a function of the shape and distribution of sources on the nucleus, as well as of the angle between the spin vector and the direction to the Sun (Jewitt 1997).  We take the value measured in 9P/Tempel 1 ($0.005 \le k_T \le 0.04$, Belton et al.~2011),  while recognizing that both larger and smaller values are possible on nuclei having other shapes and surface patterns of activity.  The active fraction, $k_A$, is widely variable among comets, with a modal value  of $k_A \sim 1\%$ (A'Hearn et al.~1995).  Lastly, we solve the sublimation energy balance equation assuming thermal equilibrium with sunlight at $r_H $= 1.6 AU to find $f_s = 7\times10^{-5}$ kg m$^{-2}$ s$^{-1}$ (and $dM/dt \sim$ 0.2 kg s$^{-1}$).  Substitution into Equation (\ref{spin2}) gives a range of timescales

\begin{equation}
\tau_s = (0.05 \text{ to } 0.5)r_e^2
\label{spin3}
\end{equation}

\noindent where $\tau_s$ is expressed in days and $r_e$ in meters.  At the $r_s$ = 20 m break-point inferred from Figure (\ref{N_vs_C}), we find $\tau_s$ = 20 - 200 days, which is comparable to the range of flight times inferred from the motions of the fragments (c.f.~Figure \ref{v_d_plot}).   In this sense, it is plausible to argue that the paucity of small fragments in  Figure (\ref{N_vs_C}) results from their prompt removal by centripetal disruption.  This process would contribute debris to the diffuse  components of 332P.  

\begin{figure}
\epsscale{1.0}
\plotone{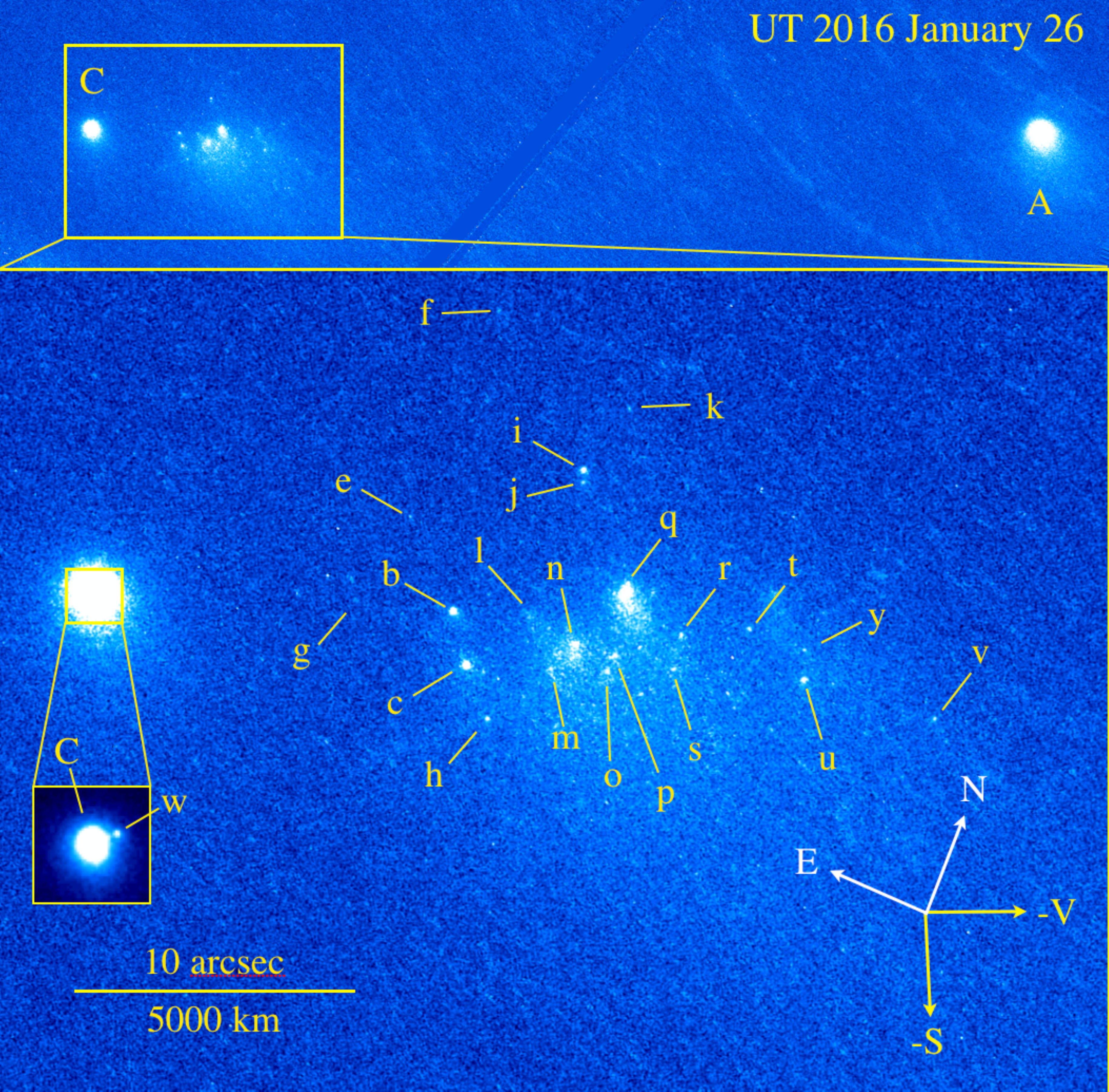}
\caption{332P on UT 2016 January 26 showing fragments measured in this work.  The image has been rotated to bring the axis of the object to the horizontal.  The wide panel at the top identifies the bright objects A and C (the parent nucleus).   A yellow box marks the region shown in the main panel, with measured fragments identified.  A further 2\arcsec~wide zoom box is included to reveal fragment $w$ in the glare of $C$.  Arrows show the cardinal directions and the projected negative velocity vector, $-V$, and the antisolar direction, $-S$, from Table (\ref{geometry}).
\label{image}}
\end{figure}

From our photometry, the parent body 332P-C has a radius $\le$275 m (mass 4.4$\times$10$^{10}$ kg, assuming density $\rho$ = 500 kg m$^{-3}$), while the sum of the volumes of all the other fragments in Figure (\ref{image}) corresponds to a sphere of radius 65 m (5.8$\times$10$^{8}$ kg).  The ratio of these masses is  $f_M \sim$1\%.  Extrapolating down to micron-sized particles using $\gamma$ = 3.6 gives a somewhat larger total cluster mass, 2.1$\times$10$^9$ kg, and a  fractional mass in the fragments of $f_M \sim$4\%. This is about 10$^3$ times larger than the reported fractional mass lost in the outburst of 2010, $f_M \sim 2\times10^{-5}$ (Ishiguro et al.~2014).  The difference is attributable in part to the much more stringent limit on the effective nucleus radius ($<$275 m vs.~$\lesssim$1.85 km) placed by the HST observations (Hui et al.~(2016) independently placed a limit of 0.5 km based on non-detections in archival data) and also to our detection of massive fragments that were not present at the time of the outburst.  With $f_M = 4\times10^{-2}$, the parent nucleus contains enough mass to sustain another $\sim$25 fragmentation events of similar size.  The Hill radius of 332P-C is about 50 km ($<$3 pixels) showing that even the closest measured fragment (w, at 200 km) is unbound.

The cause of the fragmentation in 332P, specifically, and in comets generally (e.g. Boehnhardt 2004, Fernandez 2009), remains unresolved. Ishiguro et al.~(2014) argued on the basis of the specific kinetic energy of the ejecta, and by analogy with outbursting comet 17P/Holmes (Hsieh et al.~2010, Li et al.~2011), that the 2010 photometric outburst was driven by runaway crystallization of amorphous ice.  Crystallization is exothermic, releasing up to $\sim$10$^5$ J kg$^{-1}$, and is accompanied by the release of gases formerly trapped in the intricate, sponge-like structure of amorphous ice (Notesco et al.~2003).  On the other hand, no \textit{direct} evidence for amorphous ice in comets exists, and it is not clear that  gas drag forces could be sufficient to expel fragments 10s of meters in size, as  observed, even against the low gravity of a $\le$275 m radius parent nucleus.  

\begin{figure}
\epsscale{.950}
\plotone{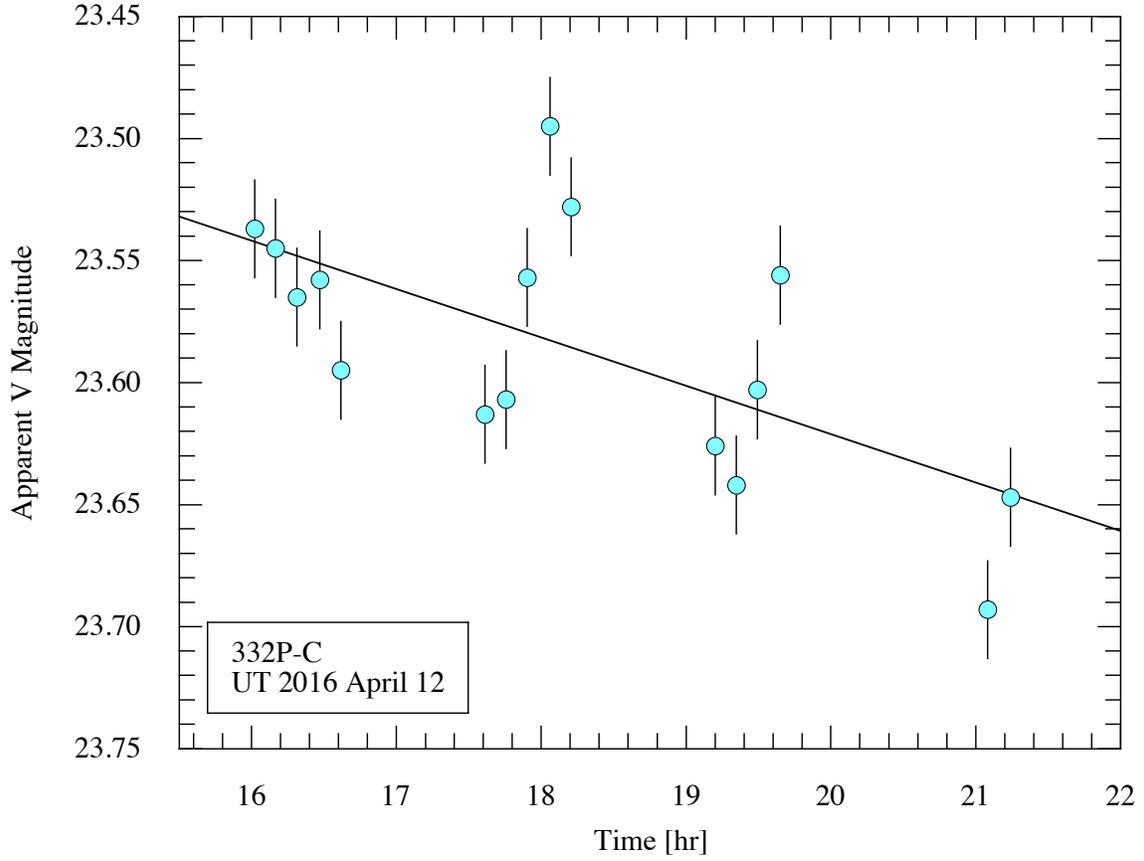}
\caption{Photometric variations in 332P-C measured on UT 2016 April 12.  Error bars of $\pm$0.02 magnitudes are shown. The solid line is a linear fit to the data to show secular fading at 0.016 magnitudes per hour.   \label{C_lightcurve}}
\end{figure}

We note that the e-folding spin-up time of  nucleus 332P-C  is $\tau_s$ = 10 to 100 years by Equation (\ref{spin3}), short enough to suggest that centripetal effects might have played a role in the ejection of fragments, in addition to the subsequent destruction of those fragments.  The lightcurve of the nucleus provides supporting evidence for this possibility.  We used a 0.2\arcsec~radius aperture with background subtraction from a 0.2\arcsec~to 0.4\arcsec~annulus to measure the brightness of component C as a function of time on UT 2016 April 12 and 13.  The results (Figure \ref{C_lightcurve}), show secular fading at about 0.016 magnitudes per day as the escape of dust from the aperture exceeds the rate of its supply.  Superimposed oscillations of the brightness  are large compared to the uncertainties of measurement and are suggestive of nucleus rotation.  Interpreted as successive puffs of dust released by the sublimation of an active patch rotating into sunlight, the effective period is near 2 hr.  Interpreted as modulation of the scattering cross-section due to rotation of an aspherical nucleus, the period would be twice this value.  Regardless, both periods are short enough to implicate rotational instability in a spherical nucleus, for which the critical period, $P = (3\pi/(G\rho))^{1/2}$, is $P$ = 4.7 hr (density $\rho$ = 500 kg m$^{-3}$, c.f.~Thomas et al.~2013, Jorda et al.~2016).   An aspherical nucleus of this density would have an even larger critical period, strengthening this conclusion.  Comet 332P  emerges as a  weakly cohesive, sub-kilometer  body probably in an excited rotational state and disintegrating over multiple orbits in response to modest heating (at 1.6 AU) by the Sun.  


\acknowledgments

We thank Pedro Lacerda for reading the manuscript and the anonymous referee for a prompt review. Based on observations made with the NASA/ESA Hubble Space Telescope, obtained  at the Space Telescope Science Institute, which is operated by the Association of Universities for Research in Astronomy, Inc., under NASA contract NAS 5-26555. These observations are associated with GO programs 14474 and 14498.  DJ appreciates support from NASA's Solar System Observations program.



{\it Facilities:}  \facility{HST (WFC3)}.




\clearpage


\clearpage

\begin{deluxetable}{clcrrrccccr}
\tabletypesize{\scriptsize}
\tablecaption{Observing Geometry 
\label{geometry}}
\tablewidth{0pt}
\tablehead{ \colhead{Visit \#} & \colhead{UT Date and Time\tablenotemark{a}} & DOY\tablenotemark{b}   & $\Delta T_p$\tablenotemark{c} & $\nu$\tablenotemark{d} & \colhead{$r_H$\tablenotemark{e}}  & \colhead{$\Delta$\tablenotemark{f}} & \colhead{$\alpha$\tablenotemark{g}}   & \colhead{$\theta_{\odot}$\tablenotemark{h}} &   \colhead{$\theta_{-v}$\tablenotemark{i}}  & \colhead{$\delta_{\oplus}$\tablenotemark{j}}   }
\startdata

1 & 2016 Jan 26 13:10 - 13:46 & 26 & -47 & 330.0 & 1.645 & 0.684 & 11.5 & 203.7 & 291.1 & -11.5 \\
2 & 2016 Jan 27 11:26 - 12:02 & 27 & -46 & 330.5 & 1.643 & 0.681 & 11.4 & 200.5 & 291.0 & -11.4 \\
3 & 2016 Jan 28 14:34 - 15:03 & 28 & -45 & 331.1 & 1.640 & 0.678 & 11.4 & 196.5 & 290.8 & -11.3 \\
4 & 2016 Apr 12 15:58 - 21:20 & 102 & 29 & 16.0 & 1.593 & 0.892 & 35.0 & 109.5 & 296.5 & 3.5		\\
5 & 2016 Apr 13 09:27 - 10:03 & 103 & 30 & 16.4 & 1.594 & 0.897 & 35.1 & 109.5 & 296.6 & 3.5 		\\
\enddata


\tablenotetext{a}{UT date and range of start times of the integrations}
\tablenotetext{b}{Day of Year, UT 2016 January 01 = 1}
\tablenotetext{c}{Number of days from perihelion (UT 2016-Mar-14 = DOY 73). Negative numbers indicate pre-perihelion observations.}
\tablenotetext{d}{True anomaly, in degrees}
\tablenotetext{e}{Heliocentric distance, in AU}
\tablenotetext{f}{Geocentric distance, in AU}
\tablenotetext{g}{Phase angle, in degrees}
\tablenotetext{h}{Position angle of the projected anti-Solar direction, in degrees}
\tablenotetext{i}{Position angle of the projected negative heliocentric velocity vector, in degrees}
\tablenotetext{j}{Angle of Earth above the orbital plane, in degrees}

\end{deluxetable}

\clearpage

\clearpage

\clearpage

\clearpage

\end{document}